\documentclass[aps,pre,superscriptaddress,twocolumn,amsmath,amssymb,showpacs]{revtex4}
\usepackage{graphicx}
\usepackage{dcolumn}
\usepackage{bm}

\begin{document}

\title{Current and universal scaling in anomalous transport}

\author{I. Goychuk}
  \affiliation{Institut f\"ur Physik,
  Universit\"at Augsburg,
  Universit\"atsstr. 1,
  D-86135 Augsburg, Germany}

\author{E. Heinsalu}
  \affiliation{Institut f\"ur Physik,
  Universit\"at Augsburg,
  Universit\"atsstr. 1,
  D-86135 Augsburg, Germany}
  \affiliation{Institute of Theoretical Physics, Tartu University,
  4 T\"ahe Street, 51010 Tartu, Estonia}
  \affiliation{Fachbereich Chemie,
  Philipps-Universit\"at Marburg, 35032 Marburg, Germany}

\author{M. Patriarca}
  \affiliation{Institut f\"ur Physik,
  Universit\"at Augsburg,
  Universit\"atsstr. 1,
  D-86135 Augsburg, Germany}

\author{G. Schmid }
  \affiliation{Institut f\"ur Physik,
  Universit\"at Augsburg,
  Universit\"atsstr. 1,
  D-86135 Augsburg, Germany}

\author{P. H\"anggi}
  \affiliation{Institut f\"ur Physik,
  Universit\"at Augsburg,
  Universit\"atsstr. 1,
  D-86135 Augsburg, Germany}

\date{\today}

\begin{abstract}
Anomalous transport in tilted periodic potentials is investigated
within the framework of the \textit{fractional} Fokker-Planck
dynamics and the underlying continuous time random walk. The
analytical solution for the stationary, anomalous current is
obtained in closed form. We derive a universal scaling law  for
anomalous diffusion occurring in tilted periodic potentials. This
scaling relation is corroborated with precise numerical studies
covering wide parameter regimes and different shapes for the
periodic potential, being either symmetric or ratchet-like ones.
\end{abstract}


\maketitle

\graphicspath{{./fig/}}  


Over recent years we witness an increasing interest in dynamical
processes that occur in systems exhibiting anomalous diffusive
behavior, possessing prominent interdisciplinary applications that
range from physics and chemistry to biology and medicine
\cite{metzler2000}. The benchmark of anomalous diffusion is the
occurrence of a mean square displacement of the form $\langle \delta
r^2 (t) \rangle \sim t^\alpha$, where $\alpha \neq 1$. Depending on
the anomalous diffusion exponent $\alpha$ the motion can either be
subdiffusive ($0 < \alpha < 1$) or superdiffusive ($\alpha
> 1$).

In the following we focus on the subdiffusive regime. Examples for
subdiffusive transport are very diverse, encompassing phenomena such
as charge carrier transport in amorphous semiconductors, nuclear
magnetic resonance, diffusion in percolative and porous systems,
transport on fractal geometries and dynamics of a bead in a
polymeric network, as well as protein conformational dynamics
\cite{metzler2000,Sokolov,Xie}. Another topic that recently gained
attention is the transport of Brownian particles in the presence of
a periodic force, relevant in Josephson junctions, rotating dipoles
in external fields, superionic conductors, charge density waves,
synchronization phenomena, diffusion on crystal surfaces, particle
separation by electrophoresis, and biophysical processes such as
intracellular transport \cite{risken,reimannR,hanggi2005}.

With this work, we present intriguing results for anomalous
diffusion and transport under the combined action of a periodically
varying spatial force and an external constant bias $F$. In
particular, we derive a closed form expression in terms of two
quadratures for the fluctuation-assisted current and its
corresponding nonlinear mobility. Furthermore, we establish a
universal scaling relation for diffusive transport which is valid in
tilted, corrugated {\em nonlinear} periodic potentials.

We start out by presenting a novel derivation of the fractional
Fokker-Planck equation  (FFPE) from a space-continuous limit of a
continuous-time random walk (CTRW). Our derivation involves nearest
neighbors jumps only; moreover, it provides new insight and
complements prior treatments in
Refs.~\cite{metzler2000,new12,barkai2000}.

\textit{Derivation of the FFPE from the CTRW.}
We study a CTRW characterized by the probability distributions
$\psi_i(\tau)$ for the residence times $\tau$, considering only
jumps between nearest-neighbor sites on a one dimensional lattice
$\{x_i\}$, with lattice period $\Delta x$. Such a CTRW is
described by a generalized master equation (GME) for the site
populations $P_i(t)$, reading \cite{Kenkre1973,Hughes,Weiss}
\begin{eqnarray} \label{GME}
\dot P_i(t) &=& \int_0^t \{K_{i-1}^{+}(t-t') P_{i-1}(t') +
K_{i+1}^{-}(t-t') P_{i+1}(t') \nonumber \\
&-& [K_i^{+}(t-t') + K_i^{-}(t-t')] P_i(t')\} \, \mathrm{d} t' \,
,
\end{eqnarray}
where the Laplace-transform of the kernel $K_i^{\pm}(t)$ is
related to the Laplace-transform of the residence time
distribution (RTD) via $\tilde K_i^{\pm}(s) = q_i^{\pm} s\tilde
\psi_i(s) / [1 - \tilde \psi_i(s)]$. The quantities $q_i^{\pm}$
are the splitting probabilities to jump from site $i$ to site $i
\pm 1$, obeying $q_i^{+} + q_i^{-} =1$.

Choosing for the RTD the Mittag-Leffler distribution,
\begin{equation} \label{ML}
\psi_i(\tau) = -\frac{\mathrm{d}}{\mathrm{d} \tau} E_{\alpha}[
-(\nu_i \tau)^{\alpha}] \, ,
\end{equation}
one obtains $\tilde K_i^{\pm}(s) = q_i^{\pm}\nu_i^{\alpha} s^{1 -
\alpha}$. Here $E_{\alpha}(z) = \sum_{n = 0}^{\infty} z^n /
\Gamma(n \alpha + 1)$ denotes the Mittag-Leffler function and the
quantity $\nu_i^{-1}$ is the time-scaling parameter at site $i$.
The corresponding GME can be recast as a fractional master
equation (FME) \cite{SokolovMetzler,GH04}, reading
\begin{eqnarray} \label{FME}
\dot P_i(t) &=& \sideset{_0}{_t} {\mathop{\hat
D}^{1-\alpha}}\{f_{i-1} \, P_{i-1}(t) + g_{i+1} \, P_{i + 1}(t)  \nonumber \\
&-& (f_i + g_i) \, P_i(t)\} \, ,
\end{eqnarray}
where the symbol $\sideset{_0}{_t}{\mathop{\hat D}^{1-\alpha}}$
stands for the integro-differential operator of the
Riemann-Liouville fractional derivative acting on a generic function
of time $\chi (t)$, as \cite{metzler2000,Sokolov,GorenfloMainardi}
\begin{equation}
\nonumber
 \sideset{_0}{_t}{\mathop{\hat D}^{1-\alpha}} \chi (t)
=\frac{1}{ \Gamma(\alpha)} \frac{\partial}{\partial t} \int_{0}^{t}
\mathrm{d} t' \, \frac{\chi(t')}{(t-t')^{1-\alpha}}\;.
\end{equation}
The quantities $f_i = q_i^{+} \nu_i^\alpha$ and $g_i = q_i^{-}
\nu_i^\alpha$ will be referred to as fractional forward and backward
rates. Using the normalization condition for the splitting
probabilities one obtains that $\nu_i = (f_i + g_i)^{1/\alpha}$, and
$q_i^+ = f_i/(f_i + g_i)$ and $q_i^- = g_i/(f_i + g_i)$, in terms of
the fractional rates.
For an arbitrary potential $U(x)$ one can set
\begin{subequations} \label{f-rate}
 \begin{align}
f_i &= (\kappa_{\alpha}/\Delta x ^2) \exp [- \beta(U_{i + 1/2} - U_i)] \, ,  \\
g_i &= (\kappa_{\alpha}/\Delta x ^2) \exp [- \beta (U_{i - 1/2} -
U_i)] \, .
\end{align}
\end{subequations}
Here $U_i \equiv U(i \Delta x)$ and $U_{i \pm 1/2} \equiv U(i \Delta x \pm \Delta x /2)$,
with $U(x)$ the total potential, $\beta = 1/ k_B T$ is the inverse of temperature,
and $\kappa_\alpha$ is the anomalous diffusion coefficient with dimension $\mathrm{cm}^2 \mathrm{s}^{-\alpha}$.
The form (\ref{f-rate}) of the
fractional rates ensures that the Boltzmann relation is satisfied,
$f_{i-1}/g_i= \exp[\beta(U_{i-1}-U_i)]$.

By use of the Laplace-transform method one can show that the FME
(\ref{FME}) can be brought into the form \cite{HilferAnton}
\begin{equation} \label{FME2}
D_{*}^{\alpha} P_i(t) \!\! = \!\! f_{i-1} \, P_{i-1}(t) + g_{i+1}
\, P_{i+1}(t) - (f_i + g_i) \, P_i(t) ,
\end{equation}
where the symbol $D_{*}^{\alpha}$ on the l.h.s. denotes the Caputo
fractional derivative \cite{GorenfloMainardi},
$$
D_{*}^{\alpha} \chi (t) = \frac{1}{\Gamma(1 - \alpha)} \int_0^t
\mathrm{d} t' \frac{1}{(t - t')^\alpha} \frac{\partial}{\partial
t'} \chi (t') \, .
$$

Let us introduce the finite difference operator $\Delta / \Delta x$,
$\Delta  P(x,t)/\Delta x = [P(x + \Delta x/2, t) - P(x - \Delta
x/2,t)]/\Delta x$, which in the limit $\Delta x \to 0$ yields the
partial derivative operator $\partial /\partial x$. Using the
fractional rates (\ref{f-rate}) the FME (\ref{FME2}) can now be
rewritten as
\begin{eqnarray} \label{FME2a}
D_{*}^{\alpha} P(x_i, t) = \kappa_{\alpha} \frac{\Delta}{\Delta x}
\left ( e^{- \beta U(x_i)} \frac{\Delta}{\Delta x} \, e^{\beta
U(x_i)} P(x_i, t) \right ) \! ,
\end{eqnarray}
where $P(x_i, t) = P(i \Delta x,t) = P_i (t) / \Delta x$. By
taking the continuous limit in Eq.~(\ref{FME2a}) one obtains the
FFPE,
\begin{eqnarray}\label{FFPE}
D_{*}^{\alpha} P(x,t)= \kappa_{\alpha} \frac{\partial }{\partial
x} \left ( e^{-\beta U(x)} \frac{\partial}{\partial x} \, e^{\beta
U(x)} P(x,t)  \right ) \, ,
\end{eqnarray}
which can  be rewritten in the well-known form with the
Riemann-Liouville fractional derivative on the r.h.s
\cite{metzler2000,new12},
\begin{eqnarray} \label{FME0}
\frac{\partial P(x, t)}{\partial t} \! = \! \kappa_{\alpha}\;
\sideset{_0}{_t}{\mathop{\hat D}^{1-\alpha}} \! \frac{\partial
}{\partial x} \left ( \! e^{-\beta U(x)} \frac{\partial}{\partial
x} e^{\beta U(x)} P(x, t) \! \right ) \!\! \, .
\end{eqnarray}

\textit{Biased diffusion.}
For a constant force $F$ the potential reads $U(x) = - F x$ and
the fractional rates (\ref{f-rate}) become site-independent, $f_i
\equiv f$ and $g_i \equiv g$, satisfying the Boltzmann relation,
i.e., $f/g = \exp (\beta F \Delta x)$ for any finite value of
$\Delta x$. Using the Laplace-transform one finds the solutions of
Eq.~(\ref{FME2}) for the mean particle position and the mean
square displacement of anomalous biased Brownian motion
\cite{metzler1998},
\begin{subequations}\label{meanXX2}
\begin{equation}
\langle x(t)\rangle = \langle x(0)\rangle + \Delta x (f - g) \,
t^{\alpha} / \Gamma(\alpha + 1) \, , \label{meanX}
\end{equation}
\begin{eqnarray}
\langle \delta x^2(t)\rangle &=& \langle \delta x^2(0)\rangle +
\Delta x ^2 (f + g) \, t^{\alpha} /\Gamma(\alpha + 1)
\label{meanX2}
\\
&+& \left (\frac{2}{\Gamma(2 \alpha + 1)} -
\frac{1}{\Gamma^2(\alpha + 1)} \right) \Delta x^2 (f - g)^2 t^{2
\alpha} \, . \nonumber
\end{eqnarray}
\end{subequations}

With respect to the case of normal diffusion the expression for
the mean square displacement contains besides a thermal
contribution $\propto t^\alpha$ also a ballistic-like term
$\propto t^{2 \alpha}$. As a consequence, a value $\alpha < 1$
does not necessarily imply subdiffusive behavior. In fact, in the
presence of bias for $1/2 < \alpha < 1$ superdiffusion takes
place.

For a finite bias $F$, the ballistic term in Eq.~(\ref{meanX2})
equals zero only in the case $\alpha = 1$, for which normal Brownian
motion is recovered. From Eqs.~(\ref{meanXX2}) one obtains then a
generalized nonlinear Einstein relation, which is nonlinear in force
and valid for a finite space step $\Delta x$,
\begin{eqnarray} \label{normal}
\frac{\langle \delta x^2(t) \rangle - \langle \delta x^2(0)
\rangle}{\langle x(t) \rangle - \langle x(0) \rangle} = \Delta x
\coth(F \beta \Delta x  / 2) \, .
\end{eqnarray}

In the limit $F \to 0$,  Eq.~(\ref{normal}) yields the well-known
Einstein relation ($\alpha = 1$), $\kappa_\alpha / \mu_\alpha (0) =
1 / \beta$, between the thermal diffusion coefficient
\begin{eqnarray} \label{kappa1}
\kappa_\alpha = \Gamma (\alpha + 1) \lim_{t \to \infty} \frac{
\langle \delta x^2(t) \rangle _{F=0} }{ 2 t^ \alpha}
\end{eqnarray}
and the linear mobility $\mu_\alpha (F=0) $, with the nonlinear
mobility $\mu_\alpha (F) =v_{\alpha} (F)/F$ being related to the
anomalous current $v_{\alpha}$ (see below), i.e.
\begin{eqnarray} \label{mobility}
\mu_\alpha (F) = \Gamma (\alpha + 1) \lim_{t \to \infty}
\frac{\langle x(t) \rangle } { F t^\alpha} \;.
\end{eqnarray}
The same Einstein relation is valid also between the submobility and
the subdiffusion coefficient for any $\alpha < 1$ \cite{new3}, as
the ballistic term in the mean square displacement (\ref{meanX2})
vanishes for $F=0$.

However, a relation analogous to the generalized nonlinear Einstein
relation (\ref{normal}) ceases to be valid for $\alpha < 1$ for any
finite $F$, as the mean square displacement becomes dominated by the
ballistic contribution in the long time limit.
Instead, from Eqs.~(\ref{meanXX2}) one obtains the following
asymptotic scaling relation,
\begin{eqnarray} \label{anomal}
\lim_{t \to \infty} \frac{\langle \delta x^2(t) \rangle}{
\langle x(t) \rangle^2} = \frac{2
\Gamma ^2(\alpha + 1)}{\Gamma(2 \alpha + 1)} - 1 \, .
\end{eqnarray}
This result no longer contains the fractional transition rates and
holds true independent of the strength of  the bias $F$ and the temperature
$T$. The relation (\ref{anomal}) has been obtained in \cite{new45}
for a continuous-time random walker that is exposed to a constant
force. As a main finding of this work we prove below that this very
relation holds true universally for
the nontrivial case of tilted nonlinearly corrugated potentials.

\textit{Tilted periodic potentials: fractional Fokker-Planck current.}
We next study the case of a periodic potential with period $L$ in
the presence of a constant force $F$. Towards this goal, it is
convenient to consider the FFPE (\ref{FFPE}), in which the Caputo
derivative appears only on the l.h.s. Analogously to the case of
normal Brownian motion \cite{risken,reimannR, stratonovich} one
obtains that the probability flux
\begin{eqnarray}
J_{\alpha}(x,t) = -\kappa_{\alpha} \, e^{-\beta U(x)}
\frac{\partial}{\partial x} \, e^{\beta U(x)} P(x,t)
\end{eqnarray}
reaches asymptotically the stationary current value, i.e.,
\begin{eqnarray} \label{exact}
D^*_{\alpha}\langle x(t) \rangle =L J_{\alpha} = v_{\alpha}(F) =
\mu_\alpha(F) F \, .
\end{eqnarray}
 The anomalous
current $v_{\alpha}(F)$ is given in closed form by
\begin{eqnarray}\label{stratSUB}
v_{\alpha}(F)  = \! \! \frac{ \kappa_{\alpha} L \, [1 - \exp(-\beta
F L)]}{\int_{0}^L \mathrm{d} x \int_{x}^{x+L} \mathrm{d} y \,
\exp(-\beta[U(x) - U(y)])} \, .
\end{eqnarray}
This result constitutes a first main result: It is the anomalous counterpart of the
current known for normal diffusion.
It obeys a form that mimics the celebrated
Stratonovich formula for normal diffusion with $\alpha=1$
\cite{stratonovich,risken,HTB90}. From Eq.~(\ref{exact}) the mean
particle position follows as
\begin{eqnarray} \label{x_per}
\langle x(t) \rangle = \langle x(0) \rangle + v_{\alpha}(F)
t^{\alpha} / \Gamma( \alpha + 1) \, .
\end{eqnarray}
\textit{Universal scaling in washboard potentials.}
We next show that the relation in (\ref{anomal}) is valid as well
for anomalous transport in  washboard-like potentials. We prove this
by mapping the dynamics onto an equivalent CTRW; i.e., we consider a
discrete state reduction of the continuous diffusion process $x(t)$:
To this aim, we introduce a lattice with sites $\{\hat x_j = j L\}$,
located at the minima of the periodic part of the potential, and
study the RTD $\hat \psi_j(\tau)$ for the hopping process between
sites $\{\hat x_j\}$. For such a system, the ratio $\hat
q_j^{+}/\hat q_j^{-}=\exp( \beta FL)$ equals that of the constant
force case, due to the choice $\Delta x = L$. Furthermore, the
analogy between the solutions (\ref{meanX}) and (\ref{x_per}), both
exhibiting an asymptotic power law $\propto t^\alpha$, implies the
same form $\hat \psi_j(\tau) \propto 1 / \tau^{1 + \alpha}$ for
$\tau \to \infty$. In fact, for $\hat \psi_j(\tau)\sim \alpha \hat
\nu_j^{-\alpha}/\Gamma(1-\alpha)\tau^{1+\alpha}$, with some suitable
scaling coefficients $\hat \nu_j$, the corresponding kernels of the
GME obey $\tilde K_j^{\pm}(s)=\hat q_j^{\pm}\hat \nu_j^\alpha
s^{1-\alpha}$ in the limit $s\to 0$. Therefore, by making use of
Tauberian theorems for the Laplace-transform \cite{Weiss}, it
follows that the asymptotic solution ($t\to\infty$) is of the form
(\ref{meanXX2}), being determined only by the asymptotic power law
behavior of the RTD \cite{MainardiFNL},  despite the fact that the
values of $\hat{f}$ and $\hat{g}$ depend on the chosen shape for the
periodic potential. Because the result in (\ref{anomal}) is
independent of $\hat{f}$ and $\hat{g}$, the scaling relation thus
still holds true. It is universal in the sense that it holds
independently of the detailed shape of the washboard potential, the
temperature $T$ and the bias strength $F$.

\begin{figure}[t]
  \includegraphics[width=7.0cm]{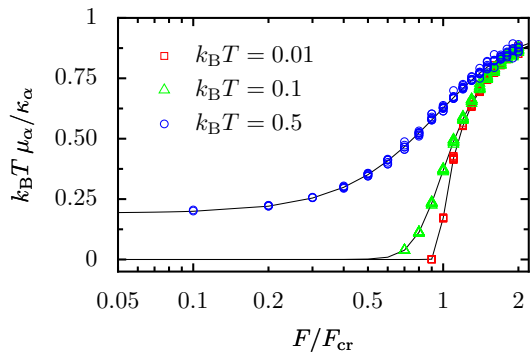}\\
  \caption{(color online). The scaled nonlinear mobility $k_{B}T \mu_\alpha(F)/\kappa_{\alpha}$
  is depicted for the case of a tilted cosine potential \textit{versus} $F/F_\mathrm{cr}$.
  The numerics for different temperatures $T$ and different $\alpha$-values, varying between $0.1-1$, fits  the analytic
  prediction (\ref{stratSUB}) (continuous lines) within the statistical errors.}
  \label{fig_a}
\end{figure}

\begin{figure}[t]
  \includegraphics[width=7.0cm]{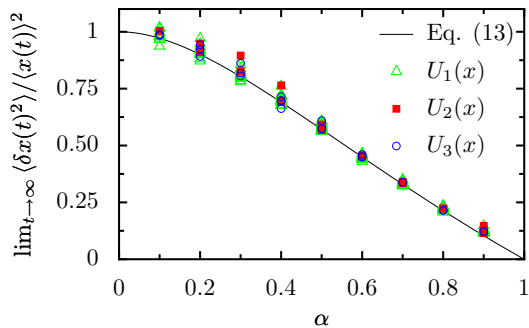}\\
  \caption{(color online). Universal scaling: Asymptotic values of the ratio $\langle \delta x^2(t) \rangle / \langle x(t) \rangle
  ^2$ as a function of the  parameter $\alpha$ for anomalous diffusion.
  All the points corresponding to the same $\alpha$ but 
  different values of $F$ and $T$
   match the function given in Eq.~(\ref{anomal}) (solid line)
  within the statistical errors. We use three different
  temperatures: $T = 0.01$ with the bias $F$ ranging between $0.9-2.0$; 
  correspondingly, $T=0.1, F=0.7-2.0$ and $T=0.5, F=0.4-2.0$.
  The open triangles correspond to the cos-potential $U_1(x)$, 
  the filled squares to the double-hump potential $U_2(x)$ and 
  the open circles
  to the asymmetric ratchet potential $U_3(x)$, see in text.
  }
\label{fig_b}
\end{figure}

\textit{Numerical verification of universal scaling.}
We have numerically tested the scaling relation (\ref{anomal}) and
the generalized Stratonovich formula (\ref{stratSUB}) through the
 simulation of the fractional CTRW in tilted washboard potentials
of various shapes and for different parameter values for $\alpha$,
$F$, and dimensionless $T$. In doing so, we not only investigate the
archetype case of a symmetric simple cosine potential $U_1(x) =
\cos(x)$,  but as well a symmetric double hump potential $U_2(x) =
\cos(x) + \cos(2x)$ and an asymmetric, ratchet-like potential
$U_3(x) = \sin(x) + \sin(2x)/3$.

In Fig.~\ref{fig_a} we depict the scaled nonlinear mobility $ k_B T
\mu_\alpha(F)/\kappa_{\alpha}$ defined by Eq.~(\ref{mobility}) for
the cosine potential. The force is in units of the critical tilt
$F_\mathrm{cr}$, which corresponds to the disappearance of potential
minima and maxima. For a given temperature $T$, all values of
$\mu_\alpha$ with $\alpha$ taken from the interval, $0.1-1$,
coincide with Eq.~(\ref{stratSUB}) (continuous lines).  For tilting
forces that exceed $F/F_{cr}=2$ the dynamics approaches the behavior
of a free CTRW being exposed to a constant bias. We further note
that the regime of linear response at low temperatures is
numerically not accessible. This is so because in this parameter
regime the corresponding escape times governing the anomalous
fluctuation-assisted transport become far too large \cite{HTB90}.

The universal scaling in tilted corrugated periodic potentials is
illustrated with Fig.~\ref{fig_b}, in which the asymptotic ratio
$\langle \delta x^2(t) \rangle / \langle x(t) \rangle ^2$ is plotted
versus $\alpha$ for the three different periodic potentials
mentioned above. For a given $\alpha$ various data are presented,
corresponding to different potential shapes and values of $F$ and
$T$. As one can deduce, these points overlap, demonstrating that the
ratio is independent of bias and temperature, as well as the
specific shape of $U(x)$. At the same time, the data fit very well
with the analytical expression (\ref{anomal}) (continuous line).

As detailed above, the long time behavior of the system is
determined only by the tail of the RTD. Since we are interested in
the asymptotic behavior ($t \to \infty$), we have used in the
numerical simulations the Pareto distribution ($0<\alpha<1$),
\begin{equation}\label{pareto}
\psi_i(\tau) = \frac{\alpha b\nu_i}{(1+b\nu_i\tau)^{1+\alpha}} =
-\frac{\mathrm{d}}{\mathrm{d}\tau}
\frac{1}{(1+b\nu_i\tau)^{\alpha}}\, .
\end{equation}
This distribution, with $b=\Gamma(1-\alpha)^{1/\alpha}$, has
precisely the same asymptotic form as  the Mittag-Leffler
distribution (\ref{ML}). For $\alpha = 1$ corresponding to a normal
Brownian process we have employed the exponential RTD $\psi_i(\tau)
=  \nu_i \exp(-\nu_i\tau)$.  In our simulations we have assumed a
space step $\Delta x \ll L$ such that $U''(x) \Delta x \ll 2 U'(x)$,
ensuring the smoothness of the periodic potential. Each trajectory
was assigned the same initial condition $x(t_0) = x_0$. Then a
residence time $\tau$ was extracted randomly from the RTD
(\ref{pareto}), time was increased to $t_1 = t_0 + \tau$ and the
particle was moved either to the right or left site with respective
probabilities $q_i^+$ and $q_i^-$. Using this procedure we  computed
the full random trajectory of the Brownian particle ($10^3$ time
steps at least). The mean displacement and the mean square
displacement were obtained as averages over $10^4$ trajectories.

\textit{Conclusions.}
We have investigated the  CTRW with power-law distributed
residence times in a  periodic potential in the presence of an
external bias. The fractional Fokker-Planck dynamics has been
derived from the corresponding space-inhomogeneous CTRW in a novel
manner. The celebrated Stratonovich solution for the stationary
current in a tilted periodic potential has been generalized to the
case of anomalous transport. Moreover, we have proven that there
exists a universal scaling law (\ref{anomal})
--- relating the mean square displacement and the mean particle
position in washboard potentials --- that does not involve the
exact form of the periodic potential, the applied bias $F$, and
the temperature $T$. This universal scaling has been verified by
numerical simulations.

Our findings for the current and the mean square fluctuations can
readily be applied to the diverse physical situations mentioned in
the introduction. The versatile and widespread use of the grand
Stratonovich result for the stationary current in case of normal
diffusion can thus readily be put to powerful use in all those
multifaceted applications where the corresponding transport
behaves anomalously.

We thank E. Barkai, J. Klafter, R. Metzler and S. Denysov for
constructive discussions. This work has been supported by the ESF
STOCHDYN project and the Estonian Science Foundation through grant
no. 5662 (EH), the DFG via research center, SFB-486, project A10,
the Volks\-wagen Foundation, via project no. I/80424.

\end{document}